\documentclass[reprint,amsmath,amssymbGaps,onecolumn,notitlepage,nofootinbib]{revtex4-1}

\usepackage{amsmath} 
\usepackage{amssymb}
\usepackage{amsfonts}
\usepackage{amstext}
\usepackage{graphicx}
\usepackage{bm}
\usepackage{color} 
\usepackage{transparent}
\usepackage{mathtools}
\usepackage{slashed}
\usepackage{hyperref}

\newcommand{\br}[1]{\left(#1\right)}
\newcommand{\sbr}[1]{\left[#1\right]}

\newcommand{\dal}{\Box \phi}
\newcommand{\dpp}{\left(\nabla \phi \right)^2}

\renewcommand{\a}{\alpha}
\renewcommand{\b}{\beta}

\newcommand{\mpl}{M_{\rm pl}}

\begin{document}

\title{Black Holes in the Scalar-Tensor Formulation of 4D Einstein-Gauss-Bonnet Gravity: \\ Uniqueness of Solutions, and a New Candidate for Dark Matter}

\author{Pedro G. S. Fernandes}
\email{p.g.s.fernandes@qmul.ac.uk}
\author{Pedro Carrilho}
% \email{p.gregoriocarrilho@qmul.ac.uk}
\author{Timothy Clifton}
% \email{t.clifton@qmul.ac.uk}
\author{David J. Mulryne}
% \email{d.mulryne@qmul.ac.uk}
\affiliation{School of Physics and Astronomy, Queen Mary University of London, Mile End Road, London, E1 4NS, UK}

%\date{\today} 

\begin{abstract}

\par In this work we study static black holes in the regularized 4D Einstein-Gauss-Bonnet theory of gravity; a shift-symmetric scalar-tensor theory that belongs to the Horndeski class. This theory features a simple black hole solution that can be written in closed form, and which we show is the unique static, spherically-symmetric and asymptotically-flat black hole vacuum solution of the theory. We further show that no asymptotically-flat, time-dependent, spherically-symmetric perturbations to this geometry are allowed, which suggests that it may be the only spherically-symmetric vacuum solution that this theory admits (a result analogous to Birkhoff's theorem). Finally, we consider the thermodynamic properties of these black holes, and find that their final state after evaporation is a remnant with a size determined by the coupling constant of the theory. We speculate that remnants of this kind from primordial black holes could act as dark matter, and we constrain the parameter space for their formation mass, as well as the coupling constant of the theory.

\end{abstract}

\maketitle

\section{Introduction} \label{intro}
\par Astonishingly, after more than a century from its inception, General Relativity (GR) remains the best known description of how gravity behaves on macroscopic scales \cite{Will:2014kxa,Ishak:2018his}, and together with quantum field theory it forms one of the two pillars of modern physics. Despite this enormous success there are, however, strong theoretical and observational reasons to believe that GR is not the final answer to our understanding of gravity, and that it may be better understood as an effective theory of something more fundamental. Strong motivation for this point of view comes from the well-known tension between Einstein's theory and quantum field theory. From a more phenomenological perspective, the apparent requirements for dark matter and dark energy also motivate consideration of modifications to Einstein's theory.

\par It is now well-known that GR is the only theory of gravity in four-dimensions that gives conserved, symmetric field equations that are no more that second-order in derivatives of the metric tensor, as proven by Lovelock \cite{Lovelock}. Thus, in order to construct gravitational theories whose field equations differ from those of GR one must relax one or more of the previous conditions \cite{Clifton:2011jh}. This leaves us with one of the following options, if we want to consider alternative theories of gravity: (i) Add extra fields that mediate the gravitational interaction, beyond just the metric tensor; (ii) Allow field equations with more than two derivatives of the metric; (iii) Work in a spacetime with dimensionality different from four; (iv) Give up on either rank-2 tensor field equations, symmetry of the field equations under exchange of indices, or divergence-free field equations; or (v) Give up on locality.

\par In this regard, one of the most well-studied classes of alternative theories of gravity are the Lovelock theories \cite{Lovelock} (see Ref. \cite{Padmanabhan:2013xyr} for a review), that fall into the third option in the previous list. The Lovelock theories of gravity are of particular interest because they are the most general theories of gravity that give covariant, conserved, second-order field equations in terms of only the metric in any arbitrary number of spacetime dimensions. In this sense, they are the most natural possible generalizations of Einstein’s theory. The first few terms in the Lovelock Lagrangian are specified by 
\begin{equation}
\label{Love3}
\mathcal{L} = \sqrt{-g} \left( - 2 \Lambda +R + \alpha \mathcal{G} + \dots \right) \, ,
\end{equation}
where the ellipsis denotes terms of higher than second power in curvature tensors, and where
\begin{equation}
\mathcal{G} \equiv R^2 - 4 R_{\mu \nu} R^{\mu \nu} + R_{\alpha \beta \mu \nu} R^{\alpha \beta \mu \nu} \,.
\label{eq:GB}
\end{equation}
It can be seen that the first two terms in this equation correspond precisely to Einstein's theory with a cosmological constant, while the third term contains the quadratic Gauss-Bonnet (GB) term $\mathcal{G}$. While in five dimensions the GB term is well-known to produce a rich generalization of Einstein's theory, in four dimensions the GB term is known to contribute precisely nothing to the field equations of the theory. This is by virtue of Chern's theorem \cite{ChernTheorem}, which shows that integrating the GB term over a four-dimensional manifold gives the constant-valued Euler characteristic.

\par Besides being the unique quadratic curvature combination appearing in the Lovelock Lagrangian, GB terms are of wide theoretical interest. String theory predicts next-to-leading order corrections at distances comparable with the string length, that are typically described by higher-order curvature terms in the action \cite{Ferrara:1996hh,Antoniadis:1997eg,Zwiebach:1985uq,Nepomechie:1985us,Callan:1986jb,Candelas:1985en,Gross:1986mw}. It is therefore of considerable interest to determine whether or not it is possible to arrive at four-dimensional theories of gravity that contain a Gauss-Bonnet term in their action, but which have non-negligible consequences at the level of the field equations. A novel proposal for just such a procedure was introduced by Glavan \& Lin\footnote{See also Ref. \cite{Tomozawa:2011gp} for an earlier but similar approach.} \cite{Glavan:2019inb}. The idea in this approach is to take the coupling parameter $\alpha$ to scale as $1/(D-4)$, and to take the limit $D \rightarrow 4$ in order to introduce a divergence that cancels the vanishing contribution of the GB term to the field equations in four dimensions (in a manner that is conceptually similar to dimensional regularization in quantum field theories). This idea has been dubbed {\it $4D$ Einstein-Gauss-Bonnet} (4DEGB) gravity, and was initially introduced in order to try and side-step Lovelock's theorem. It has attracted a great deal of attention over the past year \cite{Konoplya:2020bxa,Guo:2020zmf,Fernandes:2020rpa,Wei:2020ght,Konoplya:2020qqh,Hegde:2020xlv,Casalino:2020kbt,Ghosh:2020vpc,Doneva:2020ped,Zhang:2020qew,Ghosh:2020syx,Konoplya:2020ibi,Konoplya:2020juj,Kumar:2020owy,Kumar:2020uyz,Zhang:2020qam,HosseiniMansoori:2020yfj,Wei:2020poh,Singh:2020nwo,Churilova:2020aca,Islam:2020xmy,Mishra:2020gce,Kumar:2020xvu,Nojiri:2020tph,Singh:2020xju,Li:2020tlo,Heydari-Fard:2020sib,Konoplya:2020cbv,Jin:2020emq,Liu:2020vkh,Zhang:2020sjh,EslamPanah:2020hoj,NaveenaKumara:2020rmi,Aragon:2020qdc,Malafarina:2020pvl,Yang:2020czk,Cuyubamba:2020moe,Ying:2020bch,Shu:2020cjw,Casalino:2020pyv,Rayimbaev:2020lmz,Liu:2020evp,Zeng:2020dco,Ge:2020tid,Jusufi:2020yus,Churilova:2020mif,Kumar:2020sag,Alkac:2020zhg,Ghosh:2020cob,Yang:2020jno,Liu:2020yhu,Devi:2020uac,Jusufi:2020qyw,Konoplya:2020der,Qiao:2020hkx,Liu:2020lwc,Samart:2020sxj,Banerjee:2020stc,Narain:2020qhh,Dadhich:2020ukj,Chakraborty:2020ifg,Singh:2020mty,Banerjee:2020yhu,Narain:2020tsw,Haghani:2020ynl,Lin:2020kqe,Shaymatov:2020yte,MohseniSadjadi:2020jmc,Banerjee:2020dad,Svarc:2020fia,Hegde:2020yrd,Li:2020vpo,Wang:2020pmb,Gao:2020vhw,Zhang:2020khz,Jafarzade:2020ilt,Ghaffarnejad:2020cru,Jafarzade:2020ova,Farsam:2020pfl,Colleaux:2020wfv,Mu:2020szg,Donmez:2020rnf,Hansraj:2020rvc,Junior:2020gnu,Abdujabbarov:2020jla,MohseniSadjadi:2020qnm,Li:2020spm,Zhang:2020obn,Li:2020ozr,Lin:2021noq,Zahid:2021vdy,Kruglov:2021pdp,Liu:2021zmi,Liu:2021fzr,Zhang:2021raw,Meng:2021huz,Ding:2021iwv,Babar:2021exh,Wu:2021zyl,Donmez:2021fbk,Chen:2021gwy,Feng:2020duo,Garcia-Aspeitia:2020uwq,Wang:2021kuw,Motta:2021hvl,Kruglov:2021stm,Li:2021izh,Atamurotov:2021imh,Heydari-Fard:2021ljh,Kruglov:2021btd,Ghorai:2021uby,Ghaffarnejad:2021zbx,Zhang:2021kha,Mishra:2021qhi,Shah:2021rob,Gyulchev:2021dvt}, but has also been found to be deficient on various grounds \cite{Gurses:2020ofy,Gurses:2020rxb,Arrechea:2020evj,Arrechea:2020gjw,Bonifacio:2020vbk,Ai:2020peo,Mahapatra:2020rds,Hohmann:2020cor,Cao:2021nng}.

\par The approach of Glavan \& Lin has motivated the development of a set of alternative approaches that produce more satisfactory theories \cite{Lu:2020iav,Kobayashi:2020wqy,Mann:1992ar,Fernandes:2020nbq,Hennigar:2020lsl, Aoki:2020lig}, but which retain some of the flavour of the original idea. In this work we focus on a particular regularized 4DEGB theory that was previously obtained in Ref. \cite{Fernandes:2020nbq,Hennigar:2020lsl}, which introduces a counter-term to remove the divergent part of the theory (using a procedure introduced in 2 dimensions in Ref. \cite{Mann:1992ar}). This produces a well defined theory at the cost of introducing an additional scalar degree of freedom. Remarkably, the same theory can be obtained via a different procedure involving a regularized Kaluza-Klein reduction of the higher-dimensional EGB theory \cite{Lu:2020iav,Kobayashi:2020wqy}, by assuming a conformally invariant scalar field equation of motion \cite{Fernandes:2021dsb}, and intriguingly is the same action that appears in the context of trace anomalies \cite{Riegert:1984kt,Komargodski:2011vj}. It belongs to the Horndeski class of theories \cite{Horndeski}, much like other scalar-Gauss-Bonnet theories that exist in the literature \cite{Sotiriou:2013qea,Sotiriou:2014pfa,Saravani:2019xwx,Delgado:2020rev,Doneva:2017bvd,Silva:2017uqg,Antoniou:2017acq,Cunha:2019dwb,Collodel:2019kkx,Dima:2020yac,Herdeiro:2020wei,Berti:2020kgk,Kanti:1995vq,Kleihaus:2011tg,Kleihaus:2015aje,Cunha:2016wzk,Blazquez-Salcedo:2017txk,Nojiri:2005vv,Jiang:2013gza,Kanti:2015pda,Chakraborty:2018scm,Odintsov:2018zhw,Odintsov:2019clh,Odintsov:2020zkl,Kanti:1998jd}.

\par The regularized 4DEGB theory from Refs. \cite{Fernandes:2020nbq,Hennigar:2020lsl,Lu:2020iav,Kobayashi:2020wqy,Fernandes:2021dsb,Riegert:1984kt,Komargodski:2011vj} shares solutions with the original prescription presented by Glavan \& Lin \cite{Glavan:2019inb}, and one such solution of particular interest is that of a static and spherically-symmetric black hole. In the original proposal this solution was derived from the black hole solutions of Ref. \cite{BoulwareDeser,Wheeler:1985qd}, which were derived in the context of higher-dimensional Einstein-Gauss-Bonnet gravity. These higher-dimensional solutions have interesting properties close to the singularity, as well as an associated uniqueness theorem \cite{Cai:2001dz, Wiltshire:1985us,Wiltshire:1988uq}. It is an interesting question to determine whether or not these properties persist for the black hole solutions to the regularized 4DEGB theories. In this work we will establish results that indicate that these solutions are in fact unique in the regularized 4DEGB theory, under reasonable conditions. Furthermore, we consider the evaporation properties of these black holes, and speculate that the resultant relics might be a suitable form of dark matter.

\par This paper is structured as follows: In Section \ref{S2} we introduce the regularized 4DEGB theory, along with the respective field equations. In Section \ref{sec:uniqeness} we consider the uniqueness of the spherically-symmetric black hole solutions of the theory, and in Section \ref{sec:remnants} we speculate on the role of the resultant relics as dark matter, imposing constraints in the parameter space for their formation mass, as well as the coupling constant of the theory. We close in Section \ref{Discussion} with a discussion of our results.  We work in units such that $c=G=\hbar=k_B=1$ throughout, although in Section \ref{sec:remnants} re-introduce these constants explicitly for clarity.

\section{Regularized 4DEGB theory} \label{S2}

In this section we will introduce the scalar-tensor formulation of the regularized 4DEGB theory of Refs. \cite{Fernandes:2020nbq,Hennigar:2020lsl,Lu:2020iav,Kobayashi:2020wqy}. The action from which this theory is derived is given by
\begin{equation}
S=\frac{1}{16\pi} \int_{\mathcal{M}} d^{4} x \sqrt{-g}\Big[R+\alpha \big(4 G^{\mu \nu} \nabla_{\mu} \phi \nabla_{\nu} \phi-\phi \mathcal{G}+4 \square \phi(\nabla \phi)^{2}+2(\nabla \phi)^{4}\big) \Big] + S_m \,,
\label{eq:action}
\end{equation}
where $\mathcal{G}$ is the Gauss-Bonnet invariant defined in Eq. \eqref{eq:GB}, $\alpha$ is a coupling constant with dimensions of length squared, $\phi$ is a scalar field, and $S_{m}$ is the action associated with matter fields. This action can be obtained from the truncated Lovelock theory given in Eq. (\ref{Love3}) by putting $\Lambda$ into the matter action, and by the addition of a counter-term that consists of the Gauss-Bonnet invariant of a conformally transformed geometry $\tilde{g}_{\mu \nu}=e^{2\phi} g_{\mu \nu}$, which gives \cite{Fernandes:2020nbq,Hennigar:2020lsl}
\begin{equation}
 \lim_{D \to 4} \frac{\int_{\mathcal{M}} d^{D} x \sqrt{-g} \mathcal{G}-\int_{\mathcal{M}} d^{D} x \sqrt{-\tilde{g}} \tilde{\mathcal{G}}}{D-4} =     \int_{\mathcal{M}} d^{4} x \sqrt{-g} \br{4 G^{\mu \nu} \nabla_{\mu} \phi \nabla_{\nu} \phi-\phi \mathcal{G}+4 \square \phi(\nabla \phi)^{2}+2(\nabla \phi)^{4}} \, ,
\end{equation}
where the factor of $D-4$ in the denominator of the right-hand side occurs due to the re-scaling of the coupling parameter $\alpha\rightarrow\alpha/(D-4)$. This regularization procedure differs from the one introduced by Glavan \& Lin precisely because of the counter-term containing tildes, which removes divergences in the four-dimensional limit, and yields a well-defined scalar-tensor theory of gravity, as given in Eq. (\ref{eq:action}). In what follows we consider a positive coupling, $\alpha >0$, as supported by observational constraints \cite{Clifton:2020xhc}.

The field equations that follow from Eq~\eqref{eq:action} are obtained by varying with respect to the metric, and can be written as
\begin{equation} \label{feqs0}
    G_{\mu \nu} + \alpha \mathcal{H}_{\mu \nu}=8\pi \, T_{\mu \nu}\, ,
\end{equation}
where $T_{\mu \nu}$ is the stress-energy tensor of matter, including $\Lambda$, and where we have defined
\begin{equation} \label{feqs}
\begin{aligned}
%\mathcal{H}_{\mu\nu} =& 2G_{\mu \nu} \dpp+4P_{\mu \alpha \nu \beta}\left(\nabla^\alpha \phi \nabla^\beta \phi - \nabla^\beta \nabla^\alpha \phi \right) +4\left(\nabla_\a \phi \nabla_\mu \phi - \nabla_\alpha \nabla_\mu \phi\right) \left(\nabla^\a \phi \nabla_\nu \phi - \nabla^\a \nabla_\nu \phi\right)\\
%&+4\left(\nabla_\mu \phi \nabla_\nu \phi - \nabla_\nu \nabla_\mu \phi\right) \dal +g_{\mu \nu} \Big(2\left(\dal\right)^2 - \left( \nabla \phi\right)^4 + 2\nabla_\b \nabla_\a\phi\left(2\nabla^\a \phi \nabla^\b \phi - \nabla^\b \nabla^\a \phi \right) \Big)\,,
\mathcal{H}_{\mu\nu} =& 2G_{\mu \nu} \dpp+4P_{\mu \alpha \nu \beta}\left(\nabla^\beta \nabla^\alpha \phi - \nabla^\alpha \phi \nabla^\beta \phi\right) +4\left(\nabla_\a \phi \nabla_\mu \phi - \nabla_\alpha \nabla_\mu \phi\right) \left(\nabla^\a \phi \nabla_\nu \phi - \nabla^\a \nabla_\nu \phi\right)\\
&+4\left(\nabla_\mu \phi \nabla_\nu \phi - \nabla_\nu \nabla_\mu \phi\right) \dal +g_{\mu \nu} \Big(2\left(\dal\right)^2 - \left( \nabla \phi\right)^4 + 2\nabla_\b \nabla_\a\phi\left(2\nabla^\a \phi \nabla^\b \phi - \nabla^\b \nabla^\a \phi \right) \Big)\,,
\end{aligned}
\end{equation}
where% $P_{\alpha \beta \mu \nu} \equiv *R*_{\alpha \beta \mu \nu} = -R_{\alpha \beta \mu \nu}-g_{\alpha \nu} R_{\beta \mu}+g_{\alpha \mu} R_{\beta \nu}-g_{\beta \mu} R_{\alpha \nu}+g_{\beta \nu} R_{\alpha \mu}-\frac{1}{2}\left(g_{\alpha \mu} g_{\beta \nu}-g_{\alpha \nu} g_{\beta \mu}\right) R$
\begin{equation*}
P_{\alpha \beta \mu \nu} \equiv \frac{1}{4} \epsilon_{\alpha \beta \gamma \delta} R^{\rho \sigma \gamma \delta} \epsilon_{\rho \sigma \mu \nu} = 2\, g_{\alpha [\mu}G_{\nu] \beta} + 2\, g_{\beta [\nu} R_{\mu] \alpha} -R_{\alpha \beta \mu \nu},
\end{equation*}
is the double dual of the Riemann tensor and the square brackets denote anti-symmetrization. 

The propagation equation for $\phi$ is obtained by varying the action with respect to this field, and results in
\begin{equation} \label{scalareq}
\begin{aligned}
&R^{\mu \nu} \nabla_{\mu} \phi \nabla_{\nu} \phi - G^{\mu \nu}\nabla_\mu \nabla_\nu \phi - \dal \dpp +(\nabla_\mu \nabla_\nu \phi)^2 
- (\dal)^2 - 2\nabla_\mu \phi \nabla_\nu \phi \nabla^\mu \nabla^\nu \phi = \frac{1}{8}\mathcal{G} \, .
\end{aligned}
\end{equation}
This equation displays some interesting features. Firstly, there is a manifest conformal invariance under the transformation $g_{\mu \nu} \to g_{\mu \nu} e^{2\sigma}$ and $\phi \to \phi -\sigma$ \cite{Fernandes:2021dsb}. Secondly, it can be shown that Eq. (\ref{scalareq}) is entirely equivalent to the simple vanishing of the conformal Gauss-Bonnet invariant, $\tilde{\mathcal{G}}=0$ \cite{Fernandes:2020nbq}. Thirdly, using Eq. (\ref{scalareq}), with the trace of the field equations \eqref{feqs0}, it becomes apparent that the scalar field $\phi$ completely decouples from metric in one of the field equations, such that
\begin{equation}
    R+\frac{\alpha}{2}\mathcal{G} = -8\pi \,T,
    \label{Eq:trace}
\end{equation}
where $T=g^{\mu \nu} T_{\mu \nu}$ is the trace of the stress-energy tensor. In fact, this purely geometric relation can be shown to be a direct consequence of the conformal invariance of the scalar field equation \cite{Fernandes:2021dsb}.

We note that the theory given in Eq. (\ref{eq:action}) belongs to the Horndeski class \cite{Horndeski}, with functions $G_2=8 \alpha X^2$, $G_3=8 \alpha X$, $G_4=1+4 \alpha X$ and $G_5 = 4 \alpha \ln X$, which guarantees that the field equations (\ref{feqs0}) and (\ref{scalareq}) are at most second-order in derivatives of $g_{\mu \nu}$ and $\phi$. We further note that the action \eqref{eq:action} is shift-symmetric in the scalar field, i.e., it is invariant under the set of transformations $\phi \to \phi + \mathcal{C}$, for any constant $\mathcal{C}$. By virtue of this symmetry we acquire a Noether current with vanishing divergence \cite{Saravani:2019xwx}:
\begin{equation}
    j^{\mu} = \frac{1}{\sqrt{-g}}\frac{\delta S}{\delta (\partial_\mu \phi)}, \qquad {\rm such\,\, that} \qquad \nabla_\mu j^\mu = 0 \,. \label{eq:current}
\end{equation}
In fact, the vanishing divergence $\nabla_\mu j^\mu = 0$ implies $\partial_\mu \left(\sqrt{-g} j^\mu\right) = 0$, which also recovers the equation of motion (\ref{scalareq}). We will make use of this fact in what follows, where we will discuss the black hole solutions of this theory.

\section{black hole solutions and uniqueness}
\label{sec:uniqeness}

It is known that a vacuum solution to the field equations (\ref{feqs0}) and (\ref{scalareq}) is given by the following geometry \cite{Glavan:2019inb,Lu:2020iav}:
\begin{equation}
ds^2=-A(r) dt^2 + B(r) dr^2 + r^2\left(d\theta^2 + \sin^2 \theta d\varphi^2\right), 
\label{eq:ds2}
\end{equation}
where
\begin{equation}
B(r)^{-1}=A(r)=1+\frac{r^2}{2 \alpha} \left(1-\sqrt{1+\frac{8M\alpha}{r^3}} \right), 
\label{eq:BHsolution}
\end{equation}
and $M$ is a constant associated with mass. The corresponding scalar field profile for this solution is given up to a quadrature by
\begin{equation}
\phi'(r)=\frac{1-\sqrt{A(r)}}{r\sqrt{A(r)}},
\end{equation}
where the prime here denotes a derivative with respect to $r$. In what follows, we will show that this solution is one of only two static asymptotically-flat spherically-symmetric solutions to the regularized 4DEGB theory, and is the unique static and asymptotically flat black hole solution. We will follow this by demonstrating that the regularized 4DEGB theory admits no spherically-symmetric asymptotically-flat time-dependent perturbations to this solution, which indicates that there are no other spherically symmetric solutions (even time dependent ones) in the neighbourhood of this solution. Together, these results suggest that \eqref{eq:BHsolution} is the unique asymptotically-flat spherically-symmetric vacuum black hole solution of this theory (without assuming staticity), a result analogous to Birkhoff's theorem of GR.

\subsection{Uniqueness of Static Black Hole}
\label{sec:uniqueStat}

The first step in demonstrating the uniqueness of \eqref{eq:BHsolution} is to study the existence of solutions at spatial infinity under the assumption of asymptotic flatness. To do this we take Eq.~(\ref{eq:ds2}) as an ansatz for the most general static spherically symmetric solution, and impose asymptotic flatness by assuming that in the limit $r \to \infty$,  $A(r) \to 1$, $B(r) \to 1$ and $\phi(r) \to 0$,\footnote{Here we can make use of the scalar field shift symmetry to impose $\lim_{r \to \infty}\phi(r) = 0$.} and expand the functions of interest as a power series in $1/r$:
\begin{equation}
A(r) = 1+\sum_{n=1}^{\infty} \frac{A_{n}}{r^{n}}, \qquad B(r) = 1+\sum_{n=1}^{\infty} \frac{B_{n}}{r^{n}}, \qquad  \phi(r)=\sum_{n=1}^{\infty} \frac{q_{n}}{r^{n}}\,.
\label{eq:series}
\end{equation} 
Substituting these expressions into the the field equations \eqref{feqs0}, the (r-r) equation 
immediately tells us that $A_1=-B_1$ and the scalar field equation that $q_1=\pm B_1/2$.  
Selecting either the positive or negative branch, one finds that constants at higher order can all be fixed in 
terms of $B_1$ with no further choices, and therefore that there are two series 
solutions each of which can be written in terms 
of a single constant. We identify this constant as a mass setting $B_1 = 2 M$, 
and  at leading order one then finds the scalar charge, $q_1$, is given by
$q_1=\pm M$. Choosing $q_1=-M$ and proceeding using the field equations to fix 
coefficients order by order, one finds a series solution which matches
the Taylor expansion of the black hole solution (\ref{eq:BHsolution}) up to the order we have checked. 
On the other hand, choosing $q_1=+M$ leads to a second solution with expansion 
\begin{align}
A(r) &= 1-\frac{2M}{r}-\frac{4M^2\alpha}{r^4} + \mathcal{O}\left( \frac{1}{r^5}\right)\,,\nonumber \\ 
B(r) &= 1+\frac{2M}{r}+\frac{4M^2}{r^2} + \frac{8 M^3}{r^3}+ \frac{4 M^2(4 M^4+3 M^2 \alpha)}{r^4} +\mathcal{O}\left( \frac{1}{r^5}\right)\,, \nonumber\\ 
 \phi(r) &= \frac{M}{r} + \frac{M^2}{r^2} + \frac{4M^3}{3r^3} + \frac{2M^2(M^2+\alpha)}{r^4}  + \mathcal{O}\left( \frac{1}{r^5}\right)\,.
\label{eq:asympt}
\end{align}
We note that for this solution the expansion tells us that $B^{-1} \neq A$, which we will comment on further below.  

This analysis already indicates that there are only two static and spherically-symmetric 
asymptotically flat vacuum solutions in regularized 4DEGB, but relies on the validity of a perturbative expansion. Making use of the Noether current (\ref{eq:current}) we can go further: taking the ansatz (\ref{eq:ds2}) and utilising the expressions in Ref.~\cite{Saravani:2019xwx} we find that $j^\mu$ can be written as $j^\mu = (0, j^r,0,0)$, where
\begin{equation}
j^r = \frac{\Big(A'+2 A \phi '\Big) \left(\left(r \phi '+1\right)^2 - B\right)}{2 r^2 A B^2}.
\label{eq:j}
\end{equation}
Moreover, assuming the ansatz (\ref{eq:ds2}), Eq. \eqref{eq:current} can be integrated to give
\begin{equation}
\sqrt{AB}r^2 j^r = \mbox{constant}.
\label{eq:constcurrent}
\end{equation} 
Assuming asymptotic flatness, this constant can be seen to be zero by substituting the leading terms from either of the series solutions considered above into Eq.~\eqref{eq:j}. The same result can also be demonstrated independently of perturbation theory by integrating Eq. \eqref{eq:current} over a region of space-time that is external to the horizon, and which is bounded by the event horizon and two space-like surfaces that are identical to each other up to a translation along the Killing field $\xi^{\mu}$, which is time-like in the black hole exterior. 

To show this, we can begin by noting that Gauss' theorem means that the volume integral of $\nabla_{\mu} j^{\mu}$ can be converted to an integral of the normal component of $\vec{j}$ over the boundary. We can then see that the contribution to this integral from the integral over the event horizon will vanish. This is because the Killing vector $\xi^{\mu}$ is a generator of this horizon, and because the event horizon itself is a null surface. These two facts mean that $\xi^{\mu}$ must also be the normal to the horizon (as null vectors are normal to themselves), and therefore that the normal component of the current $\vec{j}$ must vanish on this surface (assuming that $\vec{j}$ displays the same symmetries as the spacetime, and therefore that $j^{\mu} \xi_{\mu}=0$). Now, the identical nature of the two space-like surfaces means that the integral of the normal component of $\vec{j}$ over them must sum to zero, and therefore they also contribute nothing to the integral over the boundary. We then conclude that the normal component of $\vec{j}$ must vanish at the remaining part of the boundary. This segment of the boundary is time-like, and as there is nothing special about its location, we must therefore conclude that $j^r=0$ at all points exterior to the event horizon, which demonstrates that the constant in Eq. (\ref{eq:constcurrent}) must be equal to zero.

Now, Eq.~\eqref{eq:j} allows us to calculate the two possible scalar field profiles non-perturbatively,
in terms of the functions $A$ and $B$, as 
\begin{equation}
\phi = -\frac{1}{2} \log(A) \qquad \mbox{or} \qquad \phi' = \frac{-1\pm\sqrt{B}}{r}.
\label{eq:sfprofiles}
\end{equation}
The second profile with the plus sign corresponds to the case of $q_1=-M$. 
Substituting this into the field equations, the (t-t) equation and a suitable combination of the (t-t) and (r-r) equations give us\footnote{Substitution of the second profile with the minus sign leads to the same exact field equations and solutions. In this case, however, the scalar field profile is not asymptotically flat (albeit $\phi' \to 0$, nonetheless).}
\begin{equation}
(B-1) B \left(-\alpha + B \left(\alpha-r^2\right) \right)-r(-2\alpha +B \left(2\alpha + r^2 \right))B'=0 \qquad {\rm and} \qquad (-2\alpha +B \left(r^2+\alpha \right))(A B' + A' B) = 0.
\end{equation}
The first equation admits a solution for $B$  that coincides with $B$ given by \eqref{eq:BHsolution}, while the second equation admits the solution $A =  {\cal C} B^{-1}$ where ${\cal C}$ is a constant  that can be absorbed into a redefinition of $t$ in the metric.  This scalar field profile therefore coincides with that of \eqref{eq:BHsolution} and leads to the known black hole. 

\begin{figure}[ht!]
\centering
\includegraphics[width=.5\textwidth]{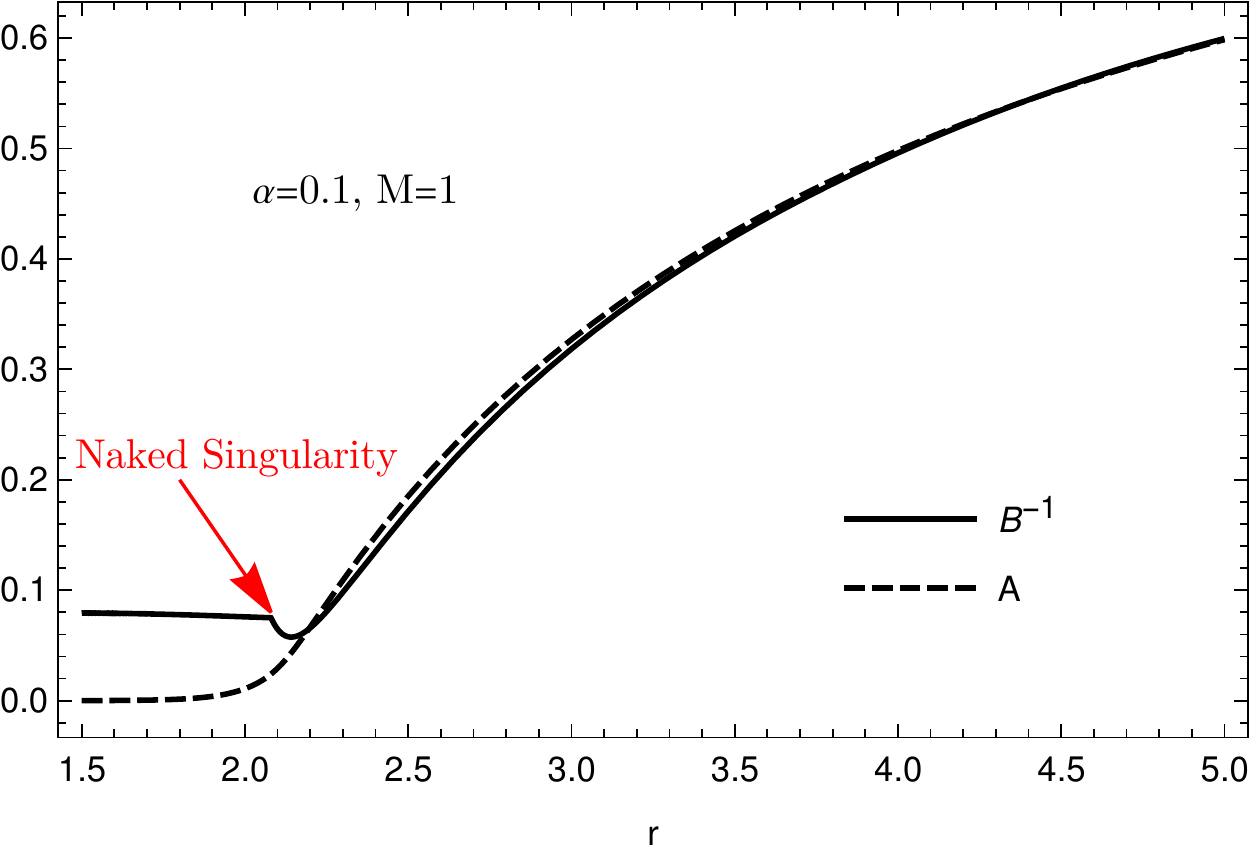}\hfill
\includegraphics[width=.5\textwidth]{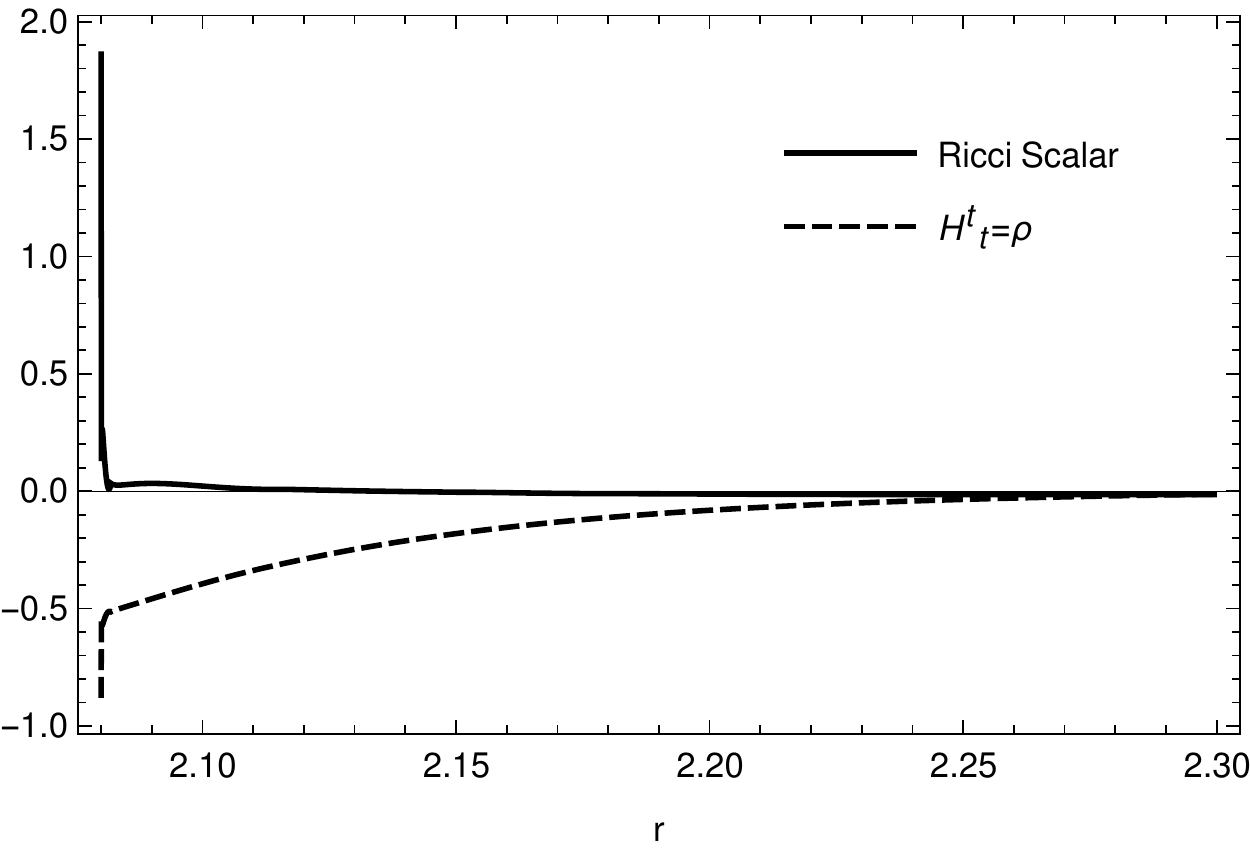}
\caption{Left: Profiles for the metric functions that solve the field equations for the scalar field profile $\phi = -\frac{1}{2} \log(A)$, for the fiducial choice $\alpha=0.1$ and $M=1$. Right: Respective Ricci scalar and (t-t) component in the field's stress energy tensor. A similar behaviour is observed for other values of the coupling $\alpha$.}
\label{fig:OtherSolution}
\end{figure}

The first scalar field profile in Eq. (\ref{eq:sfprofiles}) corresponds to the $q_1=M$ case, and the Taylor expansion of this profile matches the expansion in  Eq.~\eqref{eq:asympt}. Recall that in this case the series solution indicates that the functions $A$ and $B^{-1}$ are not equal. Studying the field equations has not allowed us to find a closed-form solution for the metric functions $A$ and $B$ in this case, so in order to make progress in understanding this solution we instead integrate the field equations numerically in $r$ from large $r$ using the series solution to provide initial conditions. As seen in Fig.~\ref{fig:OtherSolution} (left), we observe that the functions $A$ and $B^{-1}$ coincide at large $r$, but differ drastically for small values of the radial coordinate, where the function $B$ develops a kink outside of any horizon (as indicated by the arrow in the figure). As this point is approached the curvature scalar diverges, as shown in Fig. \ref{fig:OtherSolution} (right). This behaviour indicates the presence of a naked singularity. We also observe that the (t-t) component of the field's stress-energy tensor is negative, which may lead one to question whether this particular solution is of any direct physical significance at all.

\par Further evidence that demonstrates that these solutions do not represent a black hole can be obtained by expanding the metric functions in a power series around the position at which they would tend to zero, denoted $r_+$, if the solution was to describe a black hole. This gives
\begin{equation}
A = \sum_{n=1}^{\infty} a_{n} (r-r_+)^n, \qquad B^{-1} = \sum_{n=1}^{\infty} b_{n} (r-r_+)^n\,.
\label{eq:near-horizon}
\end{equation}
On substitution into the field equations, this immediately implies that $b_1 = 0$. A non-zero value of $b_1$ 
is required for this position to be the horizon of a non-extremal black hole, and hence this automatically indicates that if the solution describes a black hole, it has to be extremal. This can be seen by computing the black hole temperature $T_+$, which for the line-element of Eq. \eqref{eq:ds2}, assuming the near-horizon expansion of Eq. \eqref{eq:near-horizon}, reads \cite{Gibbons:1994ff,PhysRevD.15.2752}
\begin{equation}
    T_+ = \frac{\kappa}{2\pi} = \frac{1}{4\pi} \sbr{\frac{1}{\sqrt{-g_{tt}g_{rr}}} \frac{dg_{tt}}{dr}}_{r=r_+} = \frac{\sqrt{a_1 b_1}}{4\pi},
\end{equation}
where $\kappa$ is the surface gravity of the black hole, and $T_+$ can be seen to vanish if $b_1=0$. Moreover, a more careful analysis reveals that the aforementioned power series is incompatible with the field equations, yielding no perturbative solutions.

To summarise, we have shown that \eqref{eq:BHsolution} is the unique static spherically-symmetric and asymptotically-flat vacuum black hole solution to the regularized 4DEGB theory, and that there exists one other (likely unphysical) spherically-symmetric and asymptotically-flat solution which corresponds to a naked singularity. 

\subsection{Time-Dependent Perturbations}

Let us now generalize our considerations to allow for time dependence. To do so we will return to the ansatz \eqref{eq:ds2}, but now allow $A$ and $B$ to be functions of $t$ as well as $r$. We begin by considering spherically-symmetric time-dependent perturbations about \eqref{eq:BHsolution}. In GR such perturbations must of course be zero, by virtue of Birkhoff's theorem. We will now show that a similar result holds in regularized 4DEGB, provided we restrict our attention to spherically-symmetric, asymptotically-flat perturbations.

We denote quantities associated with the exact solution \eqref{eq:BHsolution} using a subscript $0$, and expand the metric functions as
\begin{equation}
A(t,r) = A_0 (r) + \sum_{n=1}^\infty \varepsilon^n A_n(t,r) \,,  \qquad B(t,r) = B_0 (r) + \sum_{n=1}^\infty \varepsilon^n B_n(t,r)\,,  \qquad  \phi(t,r) = \phi_0 (r) +\sum_{n=1}^\infty \varepsilon^n \phi_n(t,r)  \, ,
\label{eq:expand}
\end{equation}
where $\varepsilon$ is a small parameter. Substituting \eqref{eq:expand} into the field equations, 
and expanding to first order in $\varepsilon$, we find that the (t-r) field equation gives 
\begin{equation}
\dot {B_1}=0\,,
\end{equation} 
where the dot indicates differentiation with respect to $t$. This implies that $B_1$ must be a function of $r$ only, and by virtue of the results for the static case above we know any such function must be zero\footnote{This is because $B_1(r)$ can be re-absorbed into $B_0(r)$. It can be
explicitly verified that this is equivalent to considering a background solution with a slightly perturbed mass $M+\varepsilon\, \delta M$. New terms resultant from considering this new background solution are of $\mathcal{O}(\varepsilon)$, and can then in turn be reabsorbed into the perturbations $A_1$.}. %(as it can be reabsorbed into $B_0(r)$)
Therefore, setting $B_1=0$, we find that the (t-t) field equation is automatically 
satisfied to first order in $\varepsilon$.
On the other hand, the (r-r) equation and 
trace equation \eqref{Eq:trace} 
give 
\begin{equation}
\phi_1' = \frac{\left(2 \alpha -2 \alpha  A_0+r^2\right) \left(A_0 A_1'-A_1 A_0'\right)}{4 \alpha  A_0 \left(-r A_0'+2 A_0-2 \sqrt{A_0}\right)},
\label{eq:perturb1}
\end{equation}
and
\begin{equation}
\begin{aligned}
0=&A_1 \left(A_0' \left(\left(2 \alpha +2 \alpha  A_0+r^2\right) A_0'-4 r A_0\right)-2 A_0 \left(2 \alpha -2 \alpha  A_0+r^2\right) A_0''\right)\\&+A_0 A_1' \left(4 r A_0-\left(2 \alpha +2 \alpha  A_0+r^2\right) A_0'\right)+2 A_0{}^2 \left(2 \alpha -2 \alpha  A_0+r^2\right) A_1''\,,
\end{aligned}
\label{eq:perturb2}
\end{equation}
respectively, where the dash again indicates a derivative with respect to $r$.
Equation \eqref{eq:perturb2} has the general solution
\begin{equation}
A_1 = A_0 \left(c_1(t) + c_2(t) \int^r \frac{1}{A_0^{3/2}\left(2\alpha- 2\alpha A_0 + r^2 \right)} dr \right),
\label{eq:perturb2sol}
\end{equation}
where $c_1(t)$ and $c_2(t)$ are free functions
of time. Substituting Eq.~\eqref{eq:perturb2sol} into 
Eq.~\eqref{eq:perturb1}, the term proportional 
to $c_1(t)$ drops out, and one finds 
\begin{equation}
%\phi_1 = \int^r \frac{c_2(t)}{4 \alpha  \left(2 \left(\sqrt{A_0}-1\right) A_0-r \sqrt{A_0} A_0'\right)} dr  + c_3(t) \sim - \frac{c_2(t) r^2}{32 M \alpha} + \mathcal{O}(r) + c_3(t)\,,
\phi_1 = c_3(t) + \int^r \frac{c_2(t)}{4\alpha \sqrt{A_0} \left(2A_0-2\sqrt{A_0}-r A_0'\right)} dr  \sim c_3(t) - \frac{c_2(t) r^2}{32 M \alpha} + \mathcal{O}(r) \,,
\end{equation}
where in the last step we have made an expansion in $r$ 
near spatial infinity. 
If we now assume asymptotic flatness of the perturbations, we can set both $c_2$ and $c_3$ to zero. This implies $\phi_1=0$ and $A_1= c_1 (t) A_0$. Moreover, since $c_1$ is a function  only of $t$, this can be absorbed into a re-definition of $t$ in the line-element, such that we can effectively set $A_1=0$. With all linear perturbations 
set to zero, this further implies there are no source terms for higher-order perturbations.

We therefore conclude from this analysis that there exist no spherically-symmetric, asymptotically-flat perturbations to the solution \eqref{eq:BHsolution}, and therefore that the black hole solutions of Eq. \eqref{eq:BHsolution} are perturbatively stable. 

\section{Evaporation remnants}
\label{sec:remnants}

\par In this section we reintroduce the constants $c$, $G$, $\hbar$ and $k_B$ for clarity. Having argued for  the uniqueness of the black hole solution \eqref{eq:BHsolution},  we now turn to its further consequences.  First we observe that such a black hole has a minimum size, and that the evaporation process leads to a remnant\footnote{This is in contrast with other scalar-Gauss-Bonnet theories typically studied in the literature, where the evaporation never halts (see e.g. \cite{Kanti:1996gs}).}.

\subsection{4DEGB Black Hole Thermodynamics}

\begin{figure}[ht!]
\centering
\includegraphics[width=.6\textwidth]{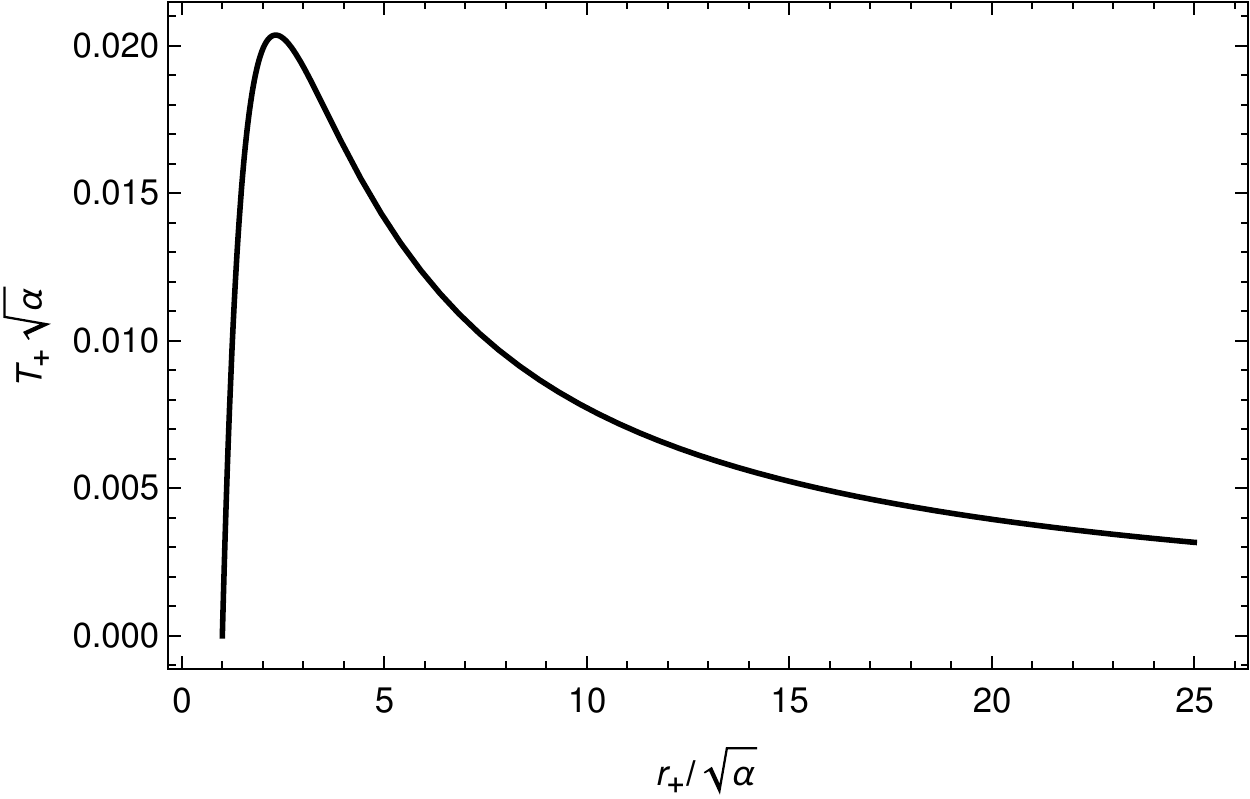}
\caption{Temperature of the 4DEGB black hole as a function of the horizon radius. One observes that as $r_+/\sqrt{\alpha} \to \sqrt{\frac{1}{2}\br{5+\sqrt{33}}} \approx 2.31782$, the temperature hits a maximum, and rapidly falls as the horizon radius decreases. The temperature approaches zero as $r_+ /\sqrt{\alpha} \to 1$.}
\label{fig:tempRH}
\end{figure}

To see that black holes leave a remnant, we first note that the black hole solution \eqref{eq:BHsolution} contains horizons located at
\begin{equation}
r_\pm = \frac{G M}{c^2} \pm \sqrt{\br{\frac{GM}{c^2}}^2-\alpha}\, ,
\end{equation}
and that $r_+$ has a minimum value of  $ r_{min} \equiv \sqrt{\alpha}$. 
The Hawking temperature of the black hole can be computed straightforwardly, giving
\begin{equation}
T_+ = \frac{\hbar c}{4\pi k_B} A'(r_+) = \frac{\hbar c}{4\pi k_B}\frac{r_+^2-\alpha }{r_+ \left(r_+^2 + 2 \alpha\right)}\,,
\end{equation}
where we observe that for $r_+=r_{min}$ the Hawking temperature vanishes, as also seen in Fig. \ref{fig:tempRH}. A similar black hole temperature profile is found in other contexts commonly related to quantum gravity, such as non-commutative models \cite{Nicolini:2005vd,Kovacik:2015yqa,Dymnikova:1992ux,Kovacik:2021qms,Gennaro:2021amf} and asymptotically safe gravity \cite{Bonanno:2006eu,Koch:2014cqa,Gennaro:2021amf}.

\begin{figure}[ht!]
\centering
\includegraphics[width=.5\textwidth]{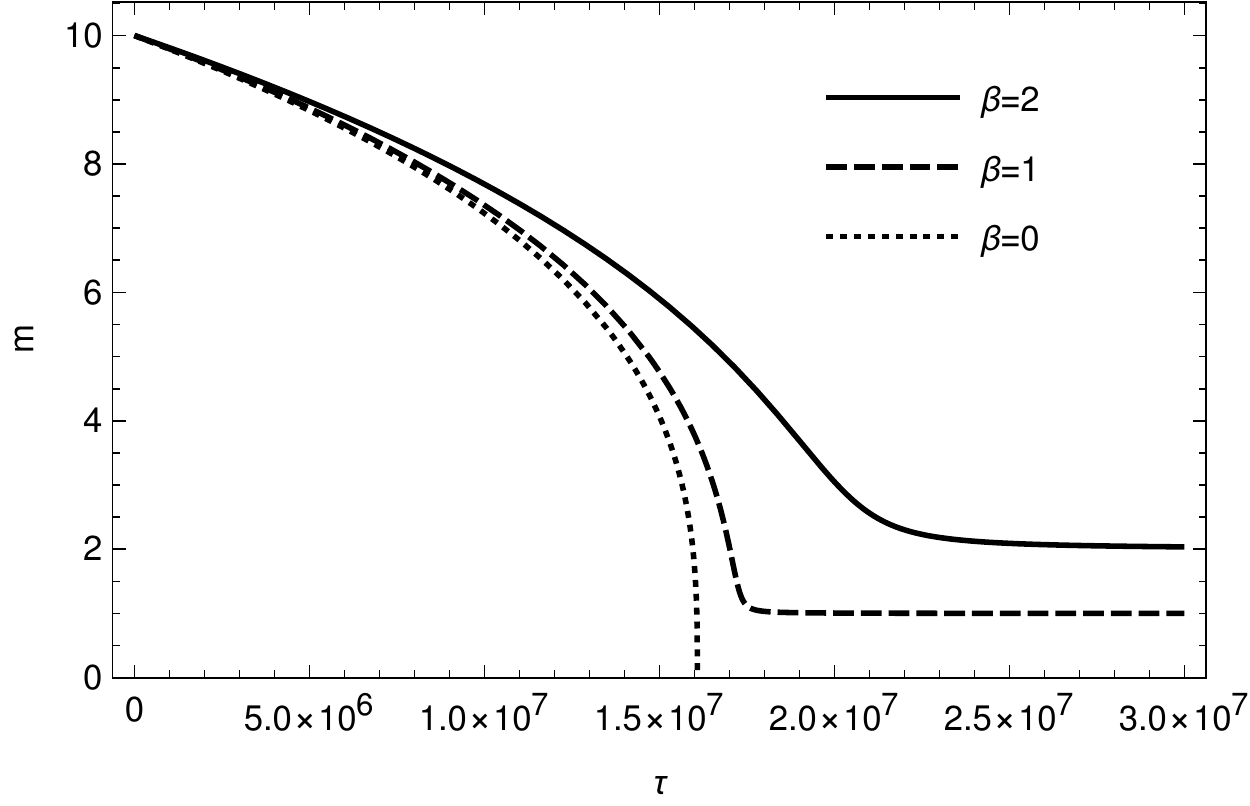}\hfill
\includegraphics[width=.5\textwidth]{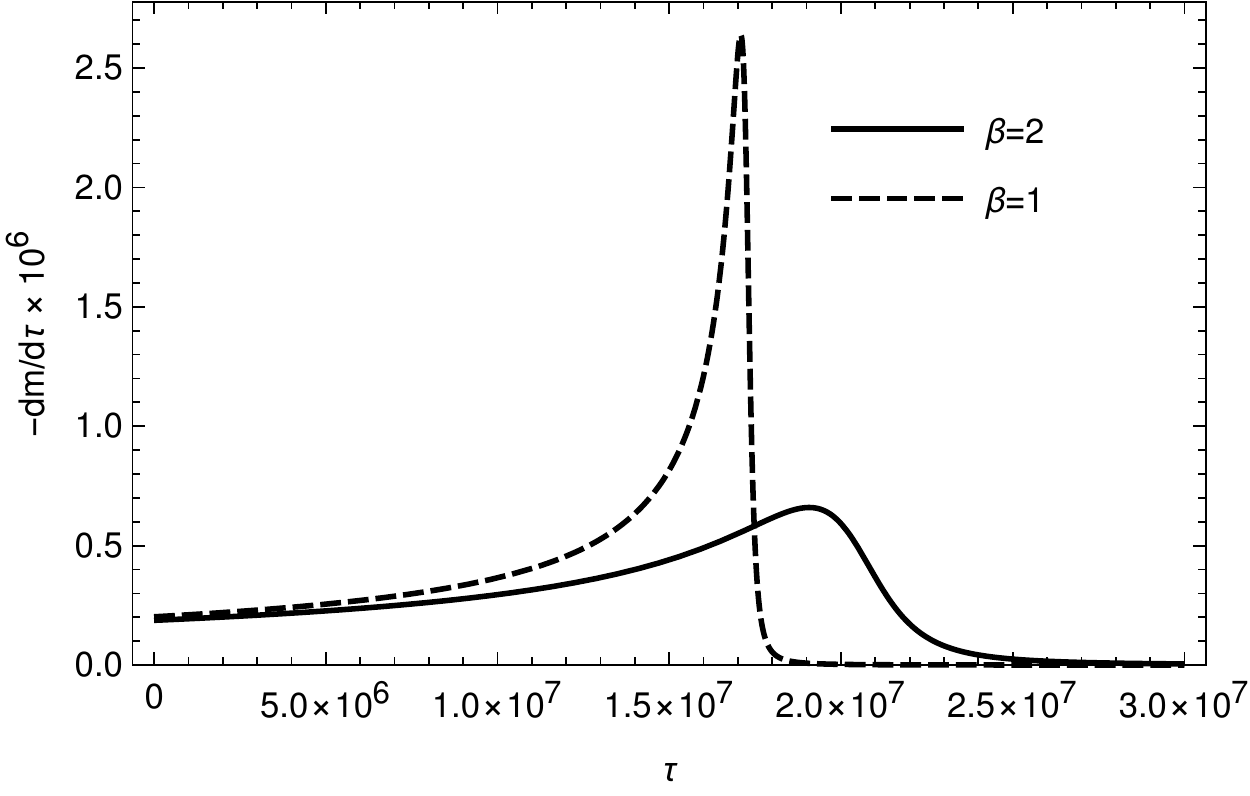}
\caption{Time evolution of the dimensionless mass $m=M/M_{pl}$ (left) and its time derivative (right) due to Hawking evaporation, for black holes with initial mass $m_0=10$ for a sample of values of the dimensionless coupling $\beta$. If $\beta=0$, black holes are described by the Schwarzschild solution of GR and evaporate completely, with $dm/d\tau \to \infty$ near the end of evaporation. 4DEGB black holes, on the other hand, approach a minimum size as $m \to \beta$ and evaporation comes to a halt.}
\label{fig:evaporation}
\end{figure}

Assuming a Stefan-Boltzmann law to estimate the mass and energy output as functions of time gives
\begin{equation}
-\frac{dE}{dt} = 4\pi r_+^2 \sigma T_+^4, \qquad {\rm and} \qquad \sigma = \frac{\pi^2 k_B^4}{60 \hbar^3 c^2}
\end{equation}
which allows us to write the following dimensionless differential equation for our black holes:
\begin{equation}
\frac{dm}{d\tau} = -\frac{1}{240\pi} \frac{\left[\beta^2 - m^2 \left(1+\sqrt{1-\frac{\beta^2}{m^2}} \right) \right]^4}{\left(1+\sqrt{1-\frac{\beta^2}{m^2}} \right)^2 m^2 \left[\beta^2 +2m^2\left(1+\sqrt{1-\frac{\beta^2}{m^2}} \right) \right]^4}\,,
\label{eq:decay}
\end{equation}
where we have defined the dimensionless quantities
\begin{equation}
\begin{aligned}
    m:=\frac{M}{M_{pl}}, \qquad \tau := \frac{t}{t_{pl}}, \qquad \beta :=\frac{\sqrt{\alpha}}{\ell_{pl}},
\end{aligned}
\label{eq:dimensionless}
\end{equation}
normalised with the Planck units
\begin{equation}
\begin{aligned}
    %M_{pl}^2:=\frac{\hbar c}{G}, \qquad \ell_{pl}^2 := \frac{\hbar G}{c^3}, \qquad t_{pl}^2 := \frac{\hbar G}{c^5}, \qquad T_{pl}^2 := \frac{\hbar c^5}{G k_B^2}.
    M_{pl}^2:=\frac{\hbar c}{G}, \qquad \ell_{pl}^2 := \frac{\hbar G}{c^3}, \qquad t_{pl}^2 := \frac{\hbar G}{c^5}.
\end{aligned}
\label{eq:planckunits}
\end{equation}
We see that the solutions to this equation will have a mass $m \to \beta$ as $\tau \to \infty$, as demonstrated by the numerical solutions displayed in Fig.~\ref{fig:evaporation}. Observe that as $m \to \beta$, $dm/d\tau \to 0$ as indicated by Eq. \eqref{eq:decay}. Furthermore, we note that the time-scale over which a black hole with dimensionless mass $m=m_0$ evaporates to its final value $m=m_f$, $t_{\mbox{ev}}$, is given by
\begin{equation}
\begin{aligned}
t_{\mbox{ev}} =& 20 \pi t_{pl}\Bigg[128 \left(m_0^3-m_f^3 + m_0^2 \sqrt{m_0^2 - \beta^2}\right) + 960\beta^2 \left(m_0 -m_f \right)+486\beta^4 \left(\frac{m_f}{m_f^2-\beta^2} - \frac{m_0}{m_0^2-\beta^2}\right)\\
&+1024 \beta^2 \sqrt{m_0^2 - \beta^2} - \frac{648 \beta^2}{\sqrt{m_0^2-\beta^2}}+729 \beta^3 \log \left(\frac{(m_0-\beta)(m_f+\beta)}{(m_0+\beta)(m_f-\beta)} \right)- \frac{8 \left(16 m_f^4+112m_f^2\beta^2-209\beta^4 \right)}{\sqrt{m_f^2-\beta^2}}\Bigg] \, .
\end{aligned}
\label{eq:evtime}
\end{equation}
This means that evaporation of these black holes in regularized 4DEGB leads to a relic, which no longer radiates, and which has a size of $\sqrt{\alpha}$. This is a favorable feature from the point of view of cosmic censorship hypothesis.

\subsection{Relic Primordial Black Holes and Dark Matter}

An immediate consequence of the end state of evaporation identified above is that the relics of black holes formed in the early universe must survive until today. Such relics may therefore contribute to the dark matter that is observed in the late Universe. The idea of primordial black holes (PBHs) contributing to the dark matter is not a new one \cite{Carr:2020gox,Carr:2020xqk}, and the possibility of Planck-size black hole relics playing the role of dark matter was first pointed out by MacGibbon \cite{MacGibbon:1987my} and has been explored by many authors \cite{Rasanen:2018fom,PhysRevD.46.645,Green:1997sz,Alexeyev:2002tg,Chen:2002tu,Chen:2004ft,Nozari:2005ah,Barrau:2003xp,Carr:1994ar,Lehmann:2019zgt,deFreitasPacheco:2020wdg,Bai:2019zcd,Kovacik:2021qms,Gennaro:2021amf,Lehmann:2021ijf} (also see Ref. \cite{Chen:2014jwq} for a review on black hole relics and their implications for the information loss paradox). In most of these studies the possible black hole relics are taken to be of Planck mass.

In the current setting there are several complications. First, the mass of the relic is now equal to $\beta \mpl \propto \sqrt{\alpha}$. Secondly, the evaporation time scale is altered, being given by Eq.~\eqref{eq:evtime}. And finally, the Friedmann equation for a flat universe in regularized 4DEGB gravity is given by\footnote{For simplicity, here we ignored a dark-radiation-like term of the form $K/a^4$, where $K$ is a free parameter. These type of terms are common in scalar-tensor theories.} \cite{Fernandes:2021dsb}
\begin{equation}
H^2 + \frac{\alpha}{c^2} H^4 = \frac{8\pi G}{3} \rho\,.
\label{eq:flatFried}
\end{equation}
The term proportional to $\alpha$ on the left-hand side of this equation may play a role in the early universe, as it scales like $H^4$. 

In what follows, we will assume that a population of black holes can form when large perturbations re-enter the horizon during the period of radiation domination after inflation ends (in  a qualitatively similar way to the process that occurs in standard general relativistic cosmology). We will further assume that 
all dark matter today consists of black hole remnants, and that the black holes initially form with a single (dimensionless) mass, $m_{\rm PBH}$. On this basis, we will estimate the allowed parameter range of $m_{\rm PBH}$ and $\beta$. Of course, it would be interesting to study further the precise details of how structure collapse and black hole formation occurs within 4DEGB, though we note that it does {\it not} appear possible to construct Oppenheimer-Snyder collapse models in regularized 4DEGB, as the scalar field from the Friedmann interior cannot be made to match that of the black hole exterior\footnote{This is true despite the fact that the first and second fundamental forms on either side of the boundary can be made to match.}. Such considerations are left to future work.

When PBHs form, their mass is given by some sizable fraction, $\gamma$, of the mass of a horizon-sized region of the universe at the time of formation. Working in units such that $\hbar=c=1$, this leads to the formula 
\begin{equation}
m_{\rm PBH} = \gamma \frac{4 \pi }{3 H_{*}^3 \mpl} \rho_*
\label{eq:PBH1}
\end{equation}
where $H_*$ is the Hubble rate at the time of re-entry, and
\begin{equation}
    \rho_*=\frac{3\mpl^2}{8\pi} \left( H_*^2 + \beta^2 \frac{H_*^4}{\mpl^2}\right)
    \label{eq:PBH2}
\end{equation}
is the density.  
Typically a value of $\gamma \sim 0.2$ is taken in the literature. The number of horizon-sized patches 
of the universe in which a black hole forms is determined 
by the amplitude and statistical properties of the large-density perturbations, and hence the fraction
of the universe's energy density that 
turns into PBHs can be taken as a free parameter.

\begin{figure}[t!]
\centering
\includegraphics[height=.35\textheight]{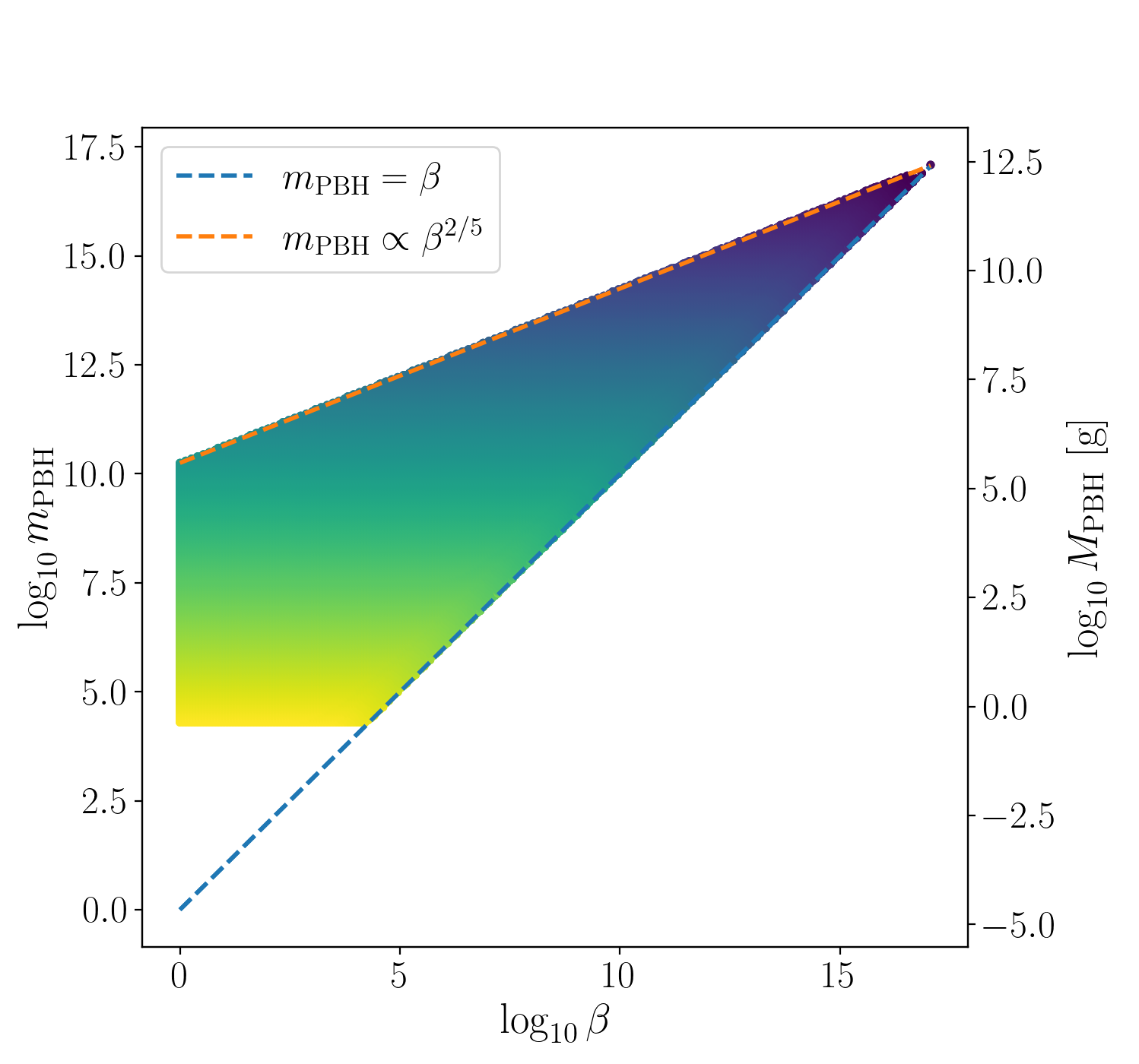}\includegraphics[height=.35\textheight]{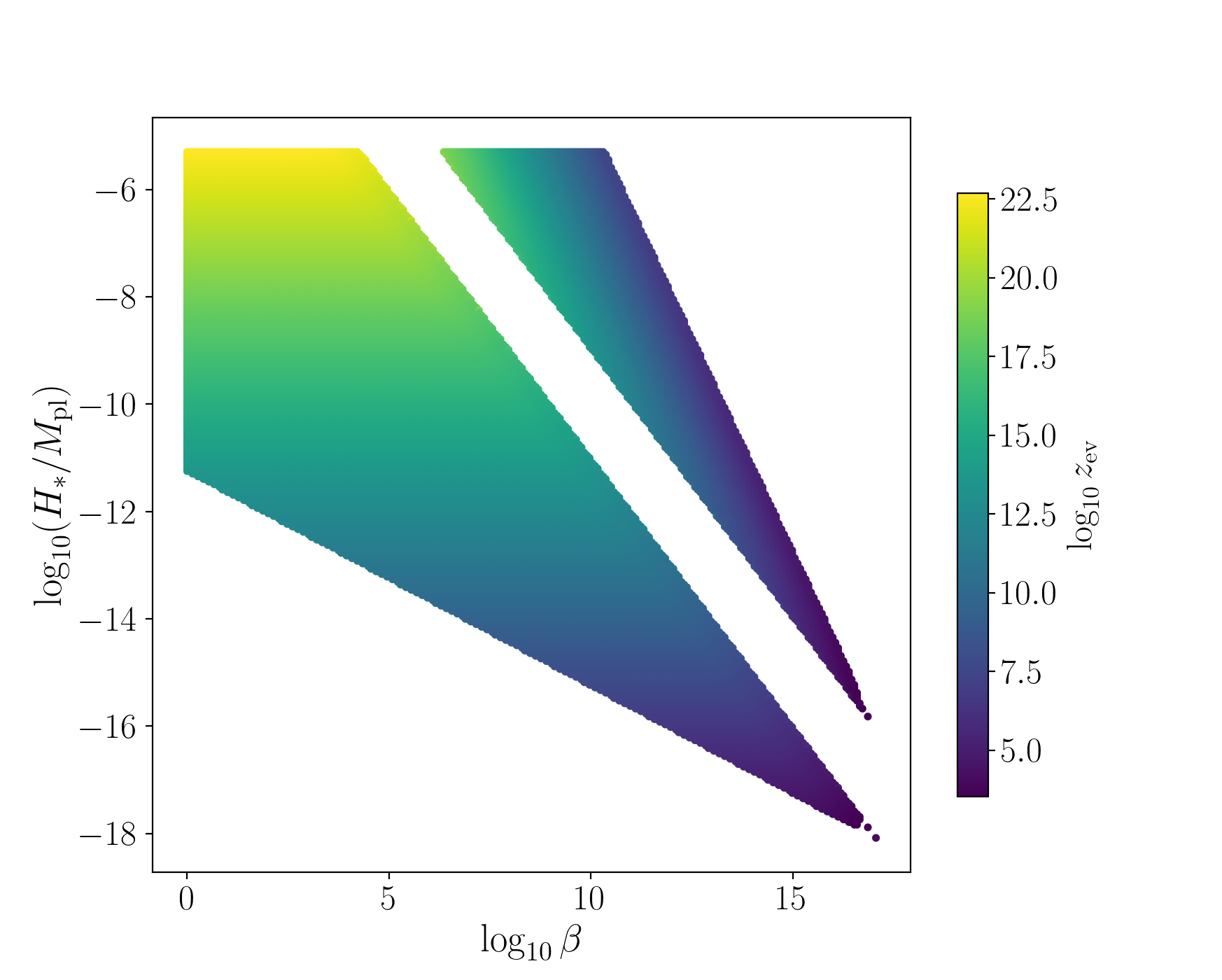}
\caption{Bounds on the mass of the PBHs (left) and Hubble rate at formation (right) as a function of $\beta$ in order that remnants make up all dark matter today. The scenario is allowed in the shaded region and colour represents the evaporation redshift, $z_{\rm ev}$.}
\label{fig:DMC}
\end{figure}

There are then two main restrictions 
on the PBH remnant dark matter scenario. 
The first is that the mass 
of the black hole 
at the time of 
formation must be greater than $\beta$.
The relations \eqref{eq:PBH1} and \eqref{eq:PBH2} given above imply 
\begin{equation}
m_{\rm PBH} = \frac{\gamma}{2} \left(\frac{\mpl}{H_* \beta} + \frac{H_* \beta}{\mpl} \right) \beta \sim 0.1 \left(\frac{\mpl}{H_* \beta} + \frac{H_* \beta}{\mpl} \right) \beta\,.
\label{eq:massPBH}
\end{equation}
 For $H_*\ll \mpl/\beta$ this formula implies $m_{\rm PBH} \propto 1/H_*$, while  for $H_*\gg \mpl/\beta$ it gives $m_{\rm PBH} \propto H_*$. For a given $\beta$ there is therefore a minimum mass of $m_{\rm PBH} \sim 0.2 \beta$ that can form, which corresponds to $H_* = \mpl/\beta$.  Since the minimum mass allowed by Eq.~\eqref{eq:massPBH} is just below the remnant mass, values of $H_*$ close to $\mpl/\beta$ are inconsistent with the outlined scenario. %, creating the gap shown in the left plot of Fig.~\ref{fig:DMC}.
 In fact, the consistency condition $m_{\rm PBH}\geq \beta$ imposes $H_* \lesssim 0.1 \mpl/\beta$ or $H_* \gtrsim 10 \mpl/\beta$.
The second main constraint is that by the time the Hubble rate reaches its value today, the density 
of dark matter and radiation must be in their correct ratio. For a given value of $\beta$, this places an upper bound on $m_{\rm PBH}$, for reasons we will explain in detail below. In turn this places an upper and lower bound on $H_*$ due to the non-linear relationship between
$m_{\rm PBH}$ and $H_*$ given above. 
The region of parameter space that satisfies both constraints is illustrated in Fig.~\ref{fig:DMC}, where the further constraint that $H_* \lesssim 5\times 10^{-6} \mpl$ required by gravitational 
wave constraints~\cite{Akrami:2018odb} has also been imposed. The colour of each point shows the time of decay of PBHs into relics in the form of the redshift $z_{\rm ev}$. We also apply the constraint $z_{\rm ev}>z_{\rm eq}\approx 3400$~\cite{Aghanim:2018eyx} to avoid relic production occurring after matter-radiation equality.

Let us now attempt to understand the 
origin of the upper
bound on $m_{\rm PBH}$. To do so we will 
assume that the 
evaporation of the black holes can 
be taken to occur instantly at some time $t_{\rm ev}$ after their formation. This time 
can be estimated using Eq.~\eqref{eq:evtime} taking $m_0 = m_{\rm PBH}$ and $m_f = 1.1 \beta$.
As $m_{\rm PBH}$ becomes larger the decay time of the black holes is pushed later into 
the universe's evolution. When PBHs evaporate, they produce radiation and this contribution to the total radiation, given by $\Delta \rho = (m_{\rm PBH}/\beta -1)\rho_{{\rm DM}}^{\rm dec}$, must be smaller than the total radiation density, which includes it. Since we are assuming relics to form all of the dark matter and the total radiation density is also well known, for sufficiently large masses, this consistency condition cannot be obeyed, otherwise the relative abundances of matter and radiation would not be correct at late time.

We can estimate the value of $m_{\rm PBH}$ at which this occurs by considering the universe today, and extrapolating into the past to 
see if a consistent evolution is possible. 
Doing so, the black hole remnants initially redshift like dust, and the ratio of remnant dark matter to radiation decreases towards the past. This behavior continues 
until the decay time is reached. At this time the dark matter density should jump by an amount given by $\Delta \rho$, and 
the radiation density, $\rho_{{\rm rad}}^{\rm dec}$ must drop by the same amount. This must occur at energy scales at least above those of matter-radiation equality for consistency with
structure formation, and hence the universe is radiation dominated at this time. Consistency then demands that $\Delta \rho < \rho_{{\rm rad}}^{\rm dec}$. The radiation density 
at the time of decay can be 
estimated by using the expression $ H^2 \approx t_{\rm ev}^{-2}/4 $, valid during radiation domination once standard cosmology is recovered. This gives a good approximation of the density, except in the fine-tuned cases where $\Delta \rho\sim \rho_{\rm rad}$. If this inequality on $\Delta \rho$ cannot be satisfied it indicates there was no consistent evolution that lead to today's energy densities, and 
employing it gives the upper bound on the mass seen in Fig.~\ref{fig:DMC}. An analytic estimate for the upper limit on the mass can be obtained by considering the GR limit of Eq.~\eqref{eq:evtime} for the evaporation time, resulting in
\begin{equation}
    \tau_{\rm ev}\approx 5120\pi m_{\rm PBH}^3\,.
\end{equation}
In addition, for $m_{\rm PBH}\gg \beta$, we can use the approximation $\Delta \rho\approx \rho_{{\rm DM}}^{\rm dec} m_{\rm PBH}/\beta$. The ratio between the densities can be calculated by relating it to matter-radiation equality (labelled by eq):
\begin{equation}
    \frac{\rho_{{\rm DM}}^{\rm dec}}{\rho_{{\rm rad}}^{\rm dec}}=\frac{a_{\rm dec}}{a_{\rm eq}}\approx\sqrt{\frac{H_{\rm eq}}{H_{\rm dec}}}=\sqrt{2H_{\rm eq}t_{\rm ev}}\,.
\end{equation}
This results in the consistency condition $\Delta \rho < \rho_{{\rm rad}}^{\rm dec}$ becoming 
\begin{equation}
   m_{\rm PBH}<(10240\pi H_{\rm eq} / M_{\rm pl})^{-1/5}\beta^{2/5}\,.
\end{equation}
This approximate limit is shown in the left plot of Fig.~\ref{fig:DMC}, in which it is clear that it works very well, with the exception of the largest values of $\beta$ allowed in this scenario for which our approximation begins to fail, as all allowed masses are very similar to $\beta$. 

In order to verify the bounds we also run more sophisticated simulations. These 
begin by fixing a value for $\beta$ and for the Hubble rate, $H_*$, at which the primordial black holes form (and hence fixing the initial energy density and black hole mass at the time of formation). The simulation then picks a value for the fraction of energy density in 
black holes at this time, taking the rest of the energy density to be radiation. Finally, the code solves the ordinary differential equations given by Eqs.~\eqref{eq:decay} and  \eqref{eq:flatFried}, assuming 
the comoving number density of radiation and black holes to be conserved. A cosmological constant of the value observed today is also included. The simulation ends when the Hubble rate reaches its observed value today, at which time the ratio between dark matter and radiation is recorded. By trying different fractions for the initial energy density of black holes for the same $\beta$ and $H_*$ the simulation can then establish if any initial fraction gives the the correct abundances today for this combination of $\beta$ and $H_*$. The simulation then picks a new $\beta$ and $H_*$ and tries again.
Our simulations also allow us to check other consistency requirements, such as the universe being radiation dominated at the time of nucleosynthesis. We find results that agree remarkably well with the simpler analytic estimate described above. 

We conclude therefore that dark matter can be generated via the mechanism of PBH evaporation in 4DEGB. We also find that remnants with a mass larger than the Planck mass (which follow when $\beta >1$) allow for the formation of PBHs at lower energy scales than in the standard scenario of Planck mass remnants. For a given energy scale, however, there is maximum value of $\beta$ above which the scenario is no longer viable, and that $\beta\sim M_{\rm pl}/H_*$ is also not permitted. The situation considered here assumed all the dark matter to be composed of relics. Should their fraction be smaller, the upper limit on PBH mass would increase in proportion with that fraction, with the corresponding limits on $H_*$ broadening too.

\section{Discussion and conclusions}
\label{Discussion}

\par Modified theories of gravity with additional degrees of freedom typically present field equations with increased complexity, and which are consequently often extremely difficult to solve in even the most symmetric situations. The regularized 4DEGB theory presented in Refs. \cite{Fernandes:2020nbq,Hennigar:2020lsl,Lu:2020iav,Kobayashi:2020wqy,Fernandes:2021dsb,Riegert:1984kt,Komargodski:2011vj}, and studied here, is an exception to this rule: it admits spherically-symmetric vacuum black hole solutions that can be written in closed form, and that can be shown to have interesting uniqueness properties. In particular we have shown that the Noether current associated with the scalar field's shift-symmetry can be used to show that the black hole solution in Eq. \eqref{eq:BHsolution} is the unique static, spherically-symmetric and asymptotically-flat vacuum solution to this theory. By further relaxing the assumption of staticity, we found that no asymptotically-vanishing time-dependent perturbations to these black hole solutions are allowed. This establishes a result only slightly less stringent than Birkhoff's theorem from GR, and suggests that the non-rotating black hole solutions of 4DEGB are perturbatively stable. 

Motivated by these results, we studied the evaporation properties of black holes in regularized 4DEGB, finding that evaporation halts at a length scale associated with the coupling constant of the theory $\alpha$, leaving relics of size $r_+ = \sqrt{\alpha}$ that no longer radiate. Assuming that a population of black holes can form, when large perturbations re-enter the horizon during the period of radiation domination after inflation ends, we have estimated the parameter range of the masses of the PBHs at formation that can constitute relic dark matter, as well as constraints on $\alpha$ that allow this. These constraints are given in Fig.~\ref{fig:DMC}, where the PBH mass at formation can range from $M \approx 0.4$~g to $M \approx 4\times10^5$~g when $\sqrt{\alpha}=\ell_{pl}$, and can reach $M \approx 2\times 10^{12}$~g when $\sqrt{\alpha} = 10^{-18} \, m$, which is the maximum value of this coupling for which this scenario is valid.

We note that the black hole geometry of Eq. \eqref{eq:BHsolution} is also present in other theoretical frameworks besides the ones mentioned before \cite{Glavan:2019inb,Lu:2020iav,Kobayashi:2020wqy,Fernandes:2020nbq,Hennigar:2020lsl, Aoki:2020lig,Fernandes:2021dsb} (consisting of the original 4DEGB procedure and well-defined regularizations thereof). Namely, this geometry also appears as a solution to the semi-classical Einstein equations when quantum corrections are considered \cite{Cai:2009ua,Cai:2014jea}. Our results on the evaporation remnants remain valid in any case. %We expect this non-trivial contribution of the Gauss-Bonnet term to have a phenomenological impact, regardless of the theoretical framework it is inserted in.

\par In conclusion, the regularized 4DEGB theory studied here has shown itself to exhibit some remarkable properties, admitting simple closed-form black hole solutions that can be shown to be unique and stable under reasonable conditions, and that leaves non-thermal relics that could contribute to the dark matter component of the Universe.

%%%%%%%%%%%%%%%%%%%%%%%%%%%  
\section*{Acknowledgements}
%%%%%%%%%%%%%%%%%%%%%%%%%%%

PF is supported by the Royal Society grant RGF/EA/180022 and acknowledges support from the project CERN/FISPAR/0027/2019. DJM is supported by a Royal Society University Research Fellowship. TC acknowledges financial support from the STFC under grant ST/P000592/1. PC acknowledges support from a UK Research and Innovation Future Leaders Fellowship (MR/S016066/1).

\bibliography{biblio}

\end{document}